%% file: main.tex
\documentclass[a4paper,11pt]{article}
\pdfoutput=1
\usepackage{jheppub}

\usepackage[utf8]{inputenc}
\usepackage[T1]{fontenc}
\usepackage{lmodern}
\usepackage{textcomp}
\usepackage{microtype}

\usepackage[english]{babel}
\usepackage[autostyle, english = american]{csquotes}
\MakeOuterQuote{"}

\usepackage{amsmath}
\usepackage{amssymb}
\usepackage{makecell} 
\usepackage{mathtools}
\usepackage{upgreek}
\usepackage{units}
\usepackage{slashed}
\usepackage{physics}
\usepackage{authblk}

\usepackage{subcaption}
\usepackage{booktabs}
\usepackage{multirow}
\usepackage{graphicx}
\usepackage{adjustbox}
\usepackage{xcolor}
\usepackage{wrapfig}
\usepackage{xspace}

\usepackage{tcolorbox}

\usepackage{rotating}
\begin{document}

\begin{titlepage}

\pagenumbering{roman}


\vspace*{1.0cm}

{\bf\boldmath\LARGE
  \begin{center}
    Prospects for Heavy Neutral Lepton Searches at Short and Medium Baseline
    Reactor Experiments \\ 

\end{center}
}

\vspace*{1cm}
\begin{center}

\vskip 2mm
N.~van Remortel$^{1}$,
M.~Colomer Molla$^{2}$,
B.~Clerbaux$^{2}$,
A.~De~Roeck$^{1,3}$, 
M.~Drewes$^{4}$,
R.~Keloth$^{5}$, 
H.~Sfar$^{1}$,
S.~Vercaemer$^{1}$,
M.~Verstraeten$^{1}$

\end{center}

\begin{center}
\bigskip
{\it 
\noindent
$^{1}$ University of Antwerp \\
$^{2}$ Université Libre de Bruxelles, Belgium\\
$^{3}$ European Organization for Nuclear Research (CERN), Geneva, Switzerland \\ 
$^{4}$ Centre for Cosmology, Particle Physics and Phenomenology, Universit\'e catholique de Louvain,  Louvain-la-Neuve, Belgium \\
$^{5}$ Vrije Universiteit Brussels, Belgium\\

}
\end{center}

\vspace*{1cm}
\begin{abstract}
  \noindent Heavy neutrinos with masses in the MeV range can in principle simultaneously explain the light neutrino masses and the origin of baryonic matter in the universe. The strongest constraints on their properties come from their potential impact on the formation of light elements in the early universe. Since these constraints rely on assumptions about the cosmic history, independent checks in the laboratory are highly desirable. In this paper, we discuss the opportunity to search for heavy neutrinos within the MeV mass range in short and medium baseline reactor neutrino experiments, using the SoLid, JUNO and TAO experiments as examples. These experiments can give the currently strongest upper bound on the mixing between the light electron neutrinos and the heavy neutrino in the 2-9 MeV mass range.
\end{abstract}
\vskip 2cm

\end{titlepage}

\pagestyle{empty}
\pagestyle{plain}
\setcounter{page}{1}
\pagenumbering{arabic}


\tableofcontents
\section{Introduction}
\label{sec:introduction}
\input{chapters/IntroductionV2}
\input{chapters/Experiments}
\label{sec:backgrounds}
\input{chapters/Background}
\section{Conclusions and outlook}
\label{sec:conclusion}
\input{chapters/Conclusion}
\bibliographystyle{JHEP}
\bibliography{references}{}
\end{document}

%% file: chapters/IntroductionV2.tex
Neutrinos are probably the most mysterious known elementary matter particles. Neutrino flavour oscillations comprise the sole established piece of empirical evidence for the existence of physics beyond the renormalisable Standard Model (SM) of particle physics that has been found in laboratory experiments on Earth.
In addition to mixing with each other, neutrinos may exhibit a mixing with new singlet fermions, often dubbed sterile neutrinos, which can provide a \emph{portal} to a hidden sector, cf.~e.g.~\cite{Alekhin:2015byh,Curtin:2018mvb,Agrawal:2021dbo} and references therein. 
The probably simplest incarnation of this idea is the extension of the SM by right-handed neutrinos; this idea is not only motivated by the observation that all other known elementary fermions have right-handed partners (and those would actually be needed for anomaly-freedom in many gauge-extensions of the SM),
but also by the fact that right-handed neutrinos appear in many popular neutrino mass models, and in particular in the type-I seesaw mechanism \cite{Minkowski:1977sc, GellMann:1980vs, Mohapatra:1979ia, Yanagida:1980xy, Schechter:1980gr, Schechter:1981cv}.
Moreover, they could potentially resolve several puzzles 
in particle physics and cosmology
that cannot be explained within the SM \cite{Drewes:2013gca,Abdullahi:2022jlv}, such as the baryon asymmetry in the Universe \cite{Canetti:2012zc} through leptogenesis \cite{Fukugita:1986hr} or the existence of dark matter
\cite{Dodelson:1993je,Shi:1998km} (cf.~\cite{Davidson:2008bu,Bodeker:2020ghk,Klaric:2021cpi} and ~\cite{Drewes:2016upu,Boyarsky:2018tvu} for reviews). 
The neutrino minimal Standard Model ($\nu$MSM) \cite{Asaka:2005an,Asaka:2005pn} represents a UV-complete example \cite{Bezrukov:2014ina} in which all of these can be achieved simultaneously with a new physics scale below the electroweak scale \cite{Canetti:2012kh,Ghiglieri:2020ulj}.

From a phenomenological point of view, sterile neutrinos represent a type of heavy neutral lepton (HNL)  
of unknown mass $m_N$. 
In the minimal scenario, their only interaction with the SM is due to their coupling to the SM weak currents \cite{Shrock:1980ct,Shrock:1981wq}, which is suppressed by the elements $|U_{\alpha N}|\ll 1$ of the complete neutrino mixing matrix (with $\alpha = e, \mu, \tau$). 
For the purpose of this work, the coupling of HNLs to weak gauge bosons ($W$ and $Z$) and Higgs bosons $h$ can be described by the phenomenological Lagrangian \cite{Atre:2009rg}:
\begin{equation}\label{eq:weak WW}
 \mathcal L
\supset
- \frac{m_W}{ v} \overline \nu_N U^*_{\alpha N} \gamma^\mu e_{L \alpha} W^+_\mu
- \frac{m_Z}{\sqrt 2 v} \overline \nu_N U^*_{\alpha N} \gamma^\mu \nu_{L \alpha} Z_\mu
- \frac{m_N}{v\sqrt{2}}  U_{\alpha N} h \overline{\nu_L}_\alpha \nu_N
+ \text{h.c.},
\end{equation}
with $m_Z$, $m_W$ the weak gauge boson masses and $v\simeq 174$ GeV the Higgs field vacuum expectation value.
The HNLs are described by four-spinors $\nu_N$ that can be Dirac or Majorana.\footnote{In realistic low scale seesaw models, the HNLs typically form so-called pseudo-Dirac fermions \cite{Shaposhnikov:2006nn,Kersten:2007vk,Moffat:2017feq} that, in some sense, lie between Dirac and Majorana \cite{Anamiati:2016uxp,Drewes:2019byd,Antusch:2023nqd}.}
The mass scale $m_N$ is unknown; while it is traditionally associated with values near the scale of Grand Unification, neutrino oscillation data can be explained even for $m_N\sim$ eV \cite{deGouvea:2005er},
and technically natural models with $m_N$ below the electroweak scale exist (cf.~e.g.~section 5 in \cite{Agrawal:2021dbo}). 

This paper explores the possibility of constraining the 
properties of HNLs with $m_N$ in the MeV range
with short and medium baseline reactor neutrino experiments, which has been poorly covered by searches to this date.
In principle, HNLs in this mass range can simultaneously explain the observed baryon asymmetry and the neutrino masses \cite{Canetti:2012kh,Domcke:2020ety}, but there are  constraints from primordial nucleosynthesis
and the subsequent history \cite{Sabti:2020yrt,Boyarsky:2020dzc,Vincent:2014rja,Diamanti:2013bia,Poulin:2016anj,Domcke:2020ety} as well as astrophysics \cite{Mastrototaro:2019vug}.
When combined with direct searches (cf.~\cite{Abdullahi:2022jlv,Antel:2023hkf}),
these essentially rule out the existence of HNLs with masses below the pion mass ($m_{\pi} = 139$~MeV) in the simplest models \cite{Drewes:2016jae,Chrzaszcz:2019inj,Bondarenko:2021cpc}, including the $\nu$MSM.
However, 
both the upper bounds from experiments\footnote{The experimental constraints can vary by orders of magnitude depending on the underlying assumptions regarding the relative size of the HNL mixing with different flavours \cite{Tastet:2021vwp}, 
which has not been fully accounted for in the global fits in the current literature \cite{Drewes:2016jae,Chrzaszcz:2019inj,Bondarenko:2021cpc}. The results obtained here are, however, comparably robust in this regard, as they do not rely on lepton number violation, and existence of other HNL decay channels has a negligible impact on the sensitivity when the decay length greatly exceeds the dimensions of the experiment.
} 
and the lower bound from big bang nucleosynthesis (BBN)\footnote{Primordial nucleosynthesis primarily constrains the HNL lifetime, cf.~\eqref{eq:HNLlifetime}, so the mixing $|U_{\alpha N}|$ with any individual SM generation $\alpha$ is in principle unconstrained from below, provided that the HNL mixes sufficiently strongly with another SM flavour $\beta\neq\alpha$. In combination with neutrino oscillation data a lower bound can be derived, but this depends on the number of HNL flavours. While strongly hierarchical $|U_{\alpha N}|^2$ are ruled out in the $\nu$MSM and models with effectively two HNL flavours \cite{Hernandez:2016kel,Drewes:2016jae,Bondarenko:2021cpc}, the constraints relax in models with more HNL flavours \cite{Chrzaszcz:2019inj,Drewes:2019mhg}. When taking this freedom into consideration, $m_N$ in the MeV range is still disfavoured by cosmology, but not strictly ruled out \cite{Domcke:2020ety}.}   
considerably relax in less minimal scenarios, and some of them can entirely be avoided 
if there is an extended dark sector \cite{deGouvea:2015euy} or new HNL interactions.\footnote{Examples for specific scenarios in which the BBN bound may be avoided include models that were designed to explain the MiniBooNE excess \cite{Fischer:2019fbw,Abdullahi:2023ejc}. While these proposals involve slightly higher HNL masses, the HNL lifetime can be kept short enough to avoid the BBN constraint for $m_N\sim 10$ MeV by lowering the new physics scale. } 
Finally, 
the cosmological constraints necessarily rely on assumptions about the early universe, while the supernova bound depends on the details of modelling the explosion \cite{Chauhan:2023sci} and has been shown to be totally avoidable in case of axion-like particles \cite{Bar:2019ifz}. 
This motivates independent direct searches for MeV scale HNLs in experiments on Earth.

In this work, we consider the model \cite{Atre:2009rg} described in equation \eqref{eq:weak WW}, with one species of Dirac-HNLs. For Majorana fermions the total decay rate 
becomes a factor 2 larger, with differences in the angular distribution \cite{Balantekin:2018ukw}. 
Since the angular distribution is randomized in the experiments considered here and their decay lengths always greatly exceed the dimensions of the experimental setup, one can in good approximation re-scale the sensitivity curves by a factor 2 for Majorana HNLs.
We further assume that the HNLs exclusively mix with the first SM generation, consistent with benchmark BC6 in \cite{Beacham:2019nyx}\footnote{While the exclusive mixing with the first generation is strictly speaking incompatible with neutrino oscillation data and strongly disfavoured in the $\nu$MSM \cite{Hernandez:2016kel,Drewes:2016jae}, it can approximately be realised in models with three flavours of HNLs \cite{Chrzaszcz:2019inj}.}.
While this does not represent a fully consistent model of neutrino masses,
it can approximately capture many aspects of the phenomenology of realistic models and provides well-defined benchmark, cf.~\cite{Drewes:2022akb}.

This paper is organised as follows. Chapter 2 describes the expected HNL signal from a reactor neutrino flux. In chapter 3, most of the experimental searches for sterile neutrinos and HNLs that have been carried out in the past years are discussed. Chapter 4 describes the opportunity to detect this signal with short-baseline reactor experiments such as SoLid and TAO, compared to the mid-baseline JUNO detector. Chapter 5 provides a discussion on the background rejection strategies that can be applied in this analysis, using the SoLid experiment as an example. Finally, conclusions are drawn in chapter 6.

%% file: chapters/Experiments.tex
\section{Signal characteristics: HNL production and decay rates}
\label{sec:signal}

If heavy neutrinos exist and couple to electron neutrinos, these HNLs must appear in nuclear $\beta$-decays due to the mixing $U_{eN}$. The HNL flux from a nuclear reactor is proportional to the initial reactor neutrino flux, $\Phi(E_{\bar{\nu}_e})$, suppressed by the coupling $\left|U_{eN}\right|^2$ and a phase space factor,
\begin{align}
\Phi(E_N) = \theta(E_N-m_N) \left|U_{eN}\right|^2 \sqrt{1-\left(\frac{m_\nu}{E_\nu}\right)^2} \Phi(E_{\bar{\nu}_e}).
\end{align}
$E_N$ denotes the energy and $m_N$ the rest mass of the massive $\bar{\nu}_N$, and $\theta(E_N-m_N)$ is the Heavyside step function, which ensures that the energy of the HNL is not smaller than its rest mass. The exponential suppression with the distance, which is controlled by the \emph{total} HNL decay rate, is neglected due to the very long decay length. This can be justified for the values of the proper HNL lifetime $\tau_N$ considered in this work, cf.~\eqref{eq:HNLlifetime} below.

The HNL flux obtained is shown in Figure~\ref{fig:HNLspectrum} as a function of the neutrino energy, for different values of the HNL mass. For illustration, the electron anti-neutrino flux from the BR2 reactor is used and is shown by the dashed line. A coupling of $\left|U_{eN}\right|^2 = 10^{-4}$ is taken as reference,  which is slightly larger than the best exclusion by the Bugey experiment of $0.8 \times 10^{-4}$~\cite{PhysRevD.52.1343}. At lower energies, the cut off from the energy conservation requirement is visible. For higher energies, the phase space factor ensures that all spectra have the same shape.

\begin{figure}[h!]
    \centering
      \includegraphics[width=0.65\textwidth]{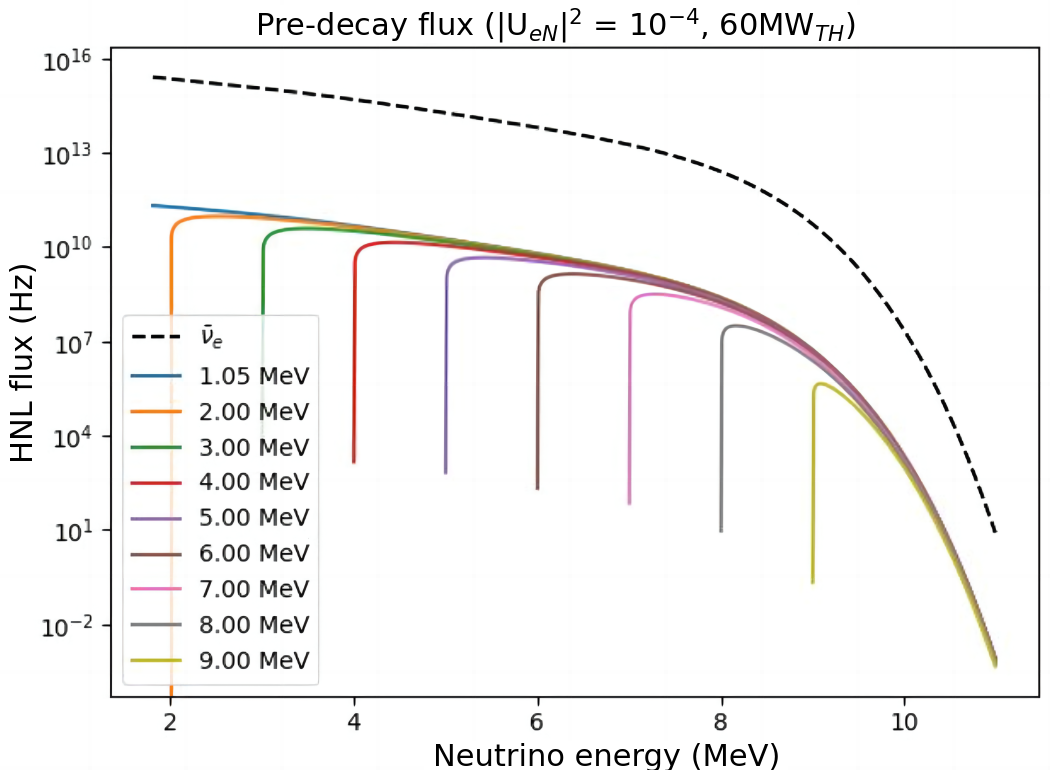}
    \caption{Expected HNL flux from the BR2 reactor in case of a coupling of $|U_{eN}|^2 = 10^{-4}$, for different proposed masses of the HNL, shown by the colors. The dotted line indicates the reactor $\bar{\nu_e}$ spectrum. \label{fig:HNLspectrum}}
\end{figure}

The HNLs will generally be unstable, albeit possibly long-lived, allowing for decays-in-flight into SM particles. In the mass range of 1 MeV - 10 MeV, the HNL can decay via three different modes.

\begin{figure}[h!]
    \centering
      \includegraphics[width=0.75\textwidth]{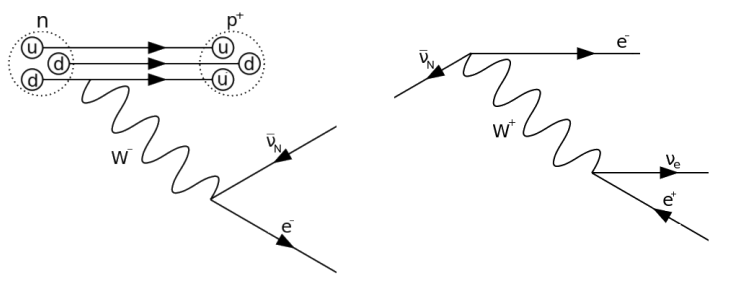}
    \caption{Feynman diagram describing the appearance of HNL after the nuclear $\beta$-decay (left) and its subsequent decay in the $e^+e^-$ mode (right). \label{fig:HNLdecaymode}}
\end{figure}

In the $e^+e^-$ mode, which can only take place if $m_N \geq 2m_e$  = 1.022 MeV, the HNL decays into a light neutrino and a $e^+e^-$ pair, as shown in Figure~\ref{fig:HNLdecaymode},
\begin{align}
\nu_N \longrightarrow \nu_j + e^+ + e^-
\end{align}

Here $\nu_j$ denotes the active neutrino, with $j$ being the electron flavor under the assumption that the HNL couples to electrons.

In the radiative mode, given that  $m_N > m(\nu_j)$, the HNL decays into a neutrino and one or two photons,
\begin{align}
\nu_N \longrightarrow\nu_j + \gamma\\
\nu_N \longrightarrow\nu_j + \gamma + \gamma
\end{align}

In the invisible mode, the HNL decays into three light neutrinos,
\begin{align}
\nu_N \longrightarrow \nu_j + \nu_k + \bar{\nu}_k
\end{align}

The precise decay rates and branching ratios for these channels are model dependent. In the energy range of reactor neutrinos (1-12 MeV), the dominant decay into visible particles is into an $e^+e^-$ pair . Indeed, the photon channel is suppressed by a factor $\sim 10^{-3}$ with respect to the $e^+e^-$ channel, and therefore can be neglected.

The $e^+e^-$ channel also happens with a much shorter decay time than the radiative decay. For reference, if $m_N$ = 5 MeV (and $|U_{eN}|^2 \sim 1$), $\tau(\nu_N \rightarrow \nu_j e^+e^-) \approx 10$ s while $\tau(\nu_N \rightarrow \nu_j\gamma) \sim 10^{10}$ s.

In this analysis, we follow the decay rates of the $\nu$MSM~\cite{Gorbunov:2007ak,Bondarenko:2018ptm,Ballett:2019bgd,Coloma:2020lgy}. The decay rate for the $e^+e^-$ mode is calculated in close analogy to the muon decay rate, taking into account the different phase-space factors due to the mass of the HNL, $m_N$. In the HNL rest frame, and assuming a Dirac HNL, one obtains 
\begin{align}\label{HNLDecayRate}
\Gamma_{N} =  \frac{G_F^2 m_N^5}{192\pi^3}|U_{eN}|^2 (1 + h(m_e^2/m_N^2) )
\end{align}
with the Fermi constant $G_F$, and $h(m_e^2/m_N^2)$ the phase-space factor calculated following Ref.~\cite{Gorbunov:2007ak}. Note that for the Majorana hypothesis, there is a factor of two difference, and a pseudo-Dirac scenario would lie in between these two.

In this work, all visible HNL decays are assumed to happen through the $e^+e^-$ mode. The $e^+e^-$ decay rate was determined as outlined in Ref.~\cite{Gorbunov:2007ak}, and the reactor spectrum was obtained from the parametrisation by Mueller~\cite{Mueller:2011nm}. The decay rates are finally evaluated as a function of the HNL mass and the interaction coupling $\left|U_{eN}\right|^2$.

\begin{figure}[h!]
    \centering
      \includegraphics[width=0.65\textwidth]{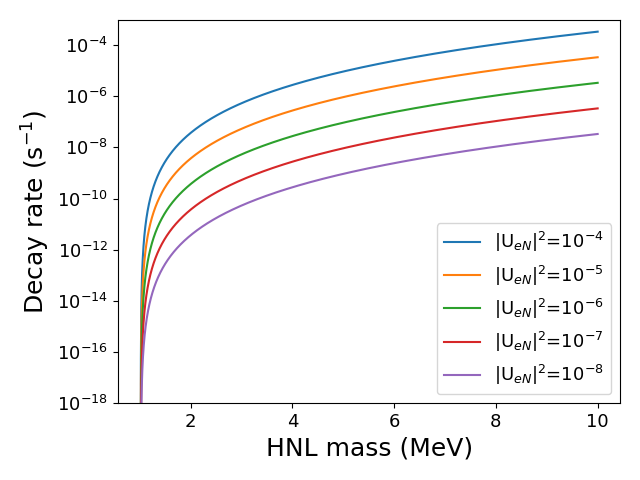}
    \caption{Decay rate into an $e^+e^-$ pair as a function of the HNL mass, for different values of the interaction coupling.
    \label{fig:HNLdecayrate}
    }
\end{figure}

The HNL lifetime in the HNL rest frame is inversely proportional to the decay rate and therefore inversely proportional to the $5^{th}$ power of the HNL mass and to the squared mixing parameter. It is given by
\begin{align}\label{eq:HNLlifetime}
  \tau_{N}^{-1} = \Gamma_N \approx 7.8 \, \text{s}^{-1} \left(\frac{m_N}{10~\text{MeV}}\right)^5 (1.4 \ |U_{eN}|^2 +  |U_{\mu N}|^2 + |U_{\tau N}|^2 )\,
\end{align}
in the mass range we consider.

\section{Summary of existing short baseline reactor experiments}
\label{sec:reactor}

Nuclear reactors are a widely used source of electron  antineutrinos in the intermediate energy range of 1 to 12~MeV. If a  mixing between electron antineutrinos and heavy neutral counterparts exists, it can produce a detectable rate of HNLs in the sensitive energy range of these experiments. Up to date, no dedicated reactor experiment exists to search for HNLs. There exist, however, a sizeable variety of reactor neutrino oscillation experiments worldwide, operating at various distances from either commercial nuclear power plants or research reactors.

For HNL masses (1--12 MeV) and couplings ($\left|U_{eN}\right|^2 < 10^{-2}$) relevant to the current best exclusion limits for a decay in flight measurement, the decay probability at distances ranging from meters to kilometers away from the production source can be considered as uniform. Therefore, the total detectable HNL flux for increasing reactor-detector distance can be compensated with a larger detector volume, and especially by a larger solid angle coverage of the detector. The HNL production rate will scale linearly with reactor power, thus favoring experiments at high power reactors. Reactor induced gamma and/or neutron backgrounds will also increase with the reactor power, but they can be further reduced by an appropriate detector location providing natural shielding or by the use of dedicated active or passive shielding.

An overview of reactor neutrino experiments capable of searching for HNL decays is given in Table \ref{Tab:NuExp}. Values are given for the sterile experiments listed in \cite{lasserre2014light} as well as for the large $\theta_{13}$ experiments (Daya Bay, Double Chooz and RENO), together with those of the experiments being the focus of this work: SoLid and the upcoming JUNO and JUNO-TAO (reffered to as "TAO"
in the following) detectors.

\begin{table}
{\small
\begin{tabular}{c | c c c c c c c}
~ &  \multicolumn{3}{c}{Reactor properties} & ~ & \multicolumn{3}{c}{Detector properties} \\
~ & Cores & \makecell{Power\\GW$_{th}$} & Fuel & \makecell{Distance\\m} & \makecell{Volume\\m$^3$} & Type & \makecell{Depth\\m.w.e.}\\
\hline
\makecell{STEREO~\cite{STEREO:2018blj}} & 1 & 0.055 & HEU & 9.4 & \makecell{1.8}& \makecell{Gd-LS} & 15\\
SoLid~\cite{SoLid:2020cen} & 1 & 0.065 & HEU & 6.2 & 1.6 & Li-PS & 8\\
NUCIFER~\cite{PhysRevD.93.112006} & 1 & 0.070 & MEU & 7.2 & 0.8 & Gd-LS & 12\\
PROSPECT~\cite{PhysRevD.103.032001} & 1 & 0.085 & HEU & 7.9 & 3.8 & Li-LS & $< 1$\\
Neutrino4~\cite{Serebrov:2017wml} & 1 & 0.090 & HEU & 6.3 & 1.8 & Gd-LS & 3-5\\
NEOS~\cite{NEOS:2016wee} & 1 & 2.73 & LEU & 23.7 & 1.0 & Gd-LS & 20\\
DANSS~\cite{Alekseev:2016llm} & 1 & 3.1 & LEU & 10.9 & 1.0 & Gd-PS & 50\\
\hline
\makecell{Double Chooz~\cite{DoubleChooz:2020pnv}} & 2 & 8.52 & LEU & 412 & \makecell{10.3} & \makecell{Gd-LS} & 120\\
\makecell{Daya Bay~\cite{DayaBay:2016ssb} (EH1)} & 2 & 4.96 & LEU & 364 & \makecell{46.8} & \makecell{Gd-LS} & 250\\
\makecell{Daya Bay (EH2)} & 4 & 10.15 & LEU & 505 & \makecell{46.8} & \makecell{Gd-LS} & 265\\
\makecell{Daya Bay (EH1+2)} & 6 & 15.11 & LEU & 458 & \makecell{93.6} & \makecell{Gd-LS} & 250-265\\
\makecell{RENO~\cite{RENO:2010vlj}} & 6 & 16.38 & LEU & 502 & \makecell{18.7} & \makecell{Gd-LS} & 110\\
\hline
\makecell{JUNO~\cite{JUNO:2021vlw}\\JUNO-TAO~\cite{JUNO:2020ijm}} & \makecell{8\\1} & \makecell{26.61\\4.6} & \makecell{LEU\\LEU} & \makecell{52.52k\\30} & \makecell{23.2k\\2.8} & \makecell{LAB\\ Gd-LS} & \makecell{1800\\25} \\
\end{tabular}
}
\caption{An overview of sterile, $\theta_{13}$ and mass hierarchy experiments which could probe the existence of HNLs. A separation is made between low (LEU), medium (MEU) and high (HEU) enriched uranium fuel in the reactor core(s). DANSS and SoLid are the only experiments in this table using plastic scintillators (PS), all others are liquid scintillator (LS) detectors.}
\label{Tab:NuExp}
\end{table}

\begin{figure}
    \centering
    \includegraphics[width=\textwidth]{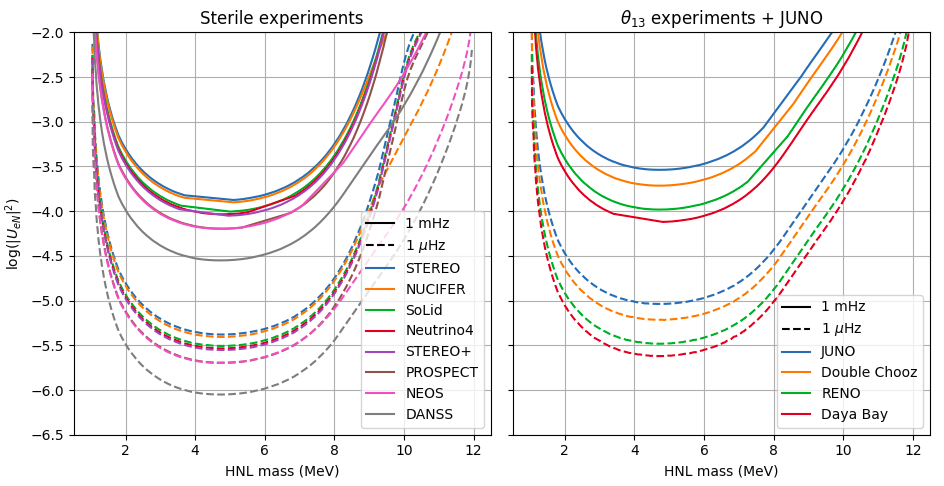}
    \caption{The coupling constants for which the expected decay rates inside the detector volume as a function of the HNL mass are, respectively 1 mHz (full lines) and $1~\mu$Hz (dashed lines). They have been evaluated for all the experiments listed in Table \ref{Tab:NuExp}.}
    \label{Fig:DecRateComp}
\end{figure}

The differences in size, distance and available reactor power roughly cancel out and provide all experiments with a comparable decay rate, as is demonstrated in Figure \ref{Fig:DecRateComp}. For each experiment, the rate at every mass-coupling pair was determined based on the detector description in Table \ref{Tab:NuExp} and the neutrino spectra parametrisation from Ref.~\cite{Mueller:2011nm}. The fission fractions have been estimated separately for each kind of reactor. For all LEU reactors, the fission fractions obtained from Table 9 in Ref.~\cite{DayaBay:2016ssb} have been used; for NUCIFER, the only experiment at a MEU reactor (Osiris), the mean values of Figure 15 in Ref.~\cite{PhysRevD.93.112006} were taken. For all experiments at HEU reactors, the internal SoLid fission fractions have been considered with 0.01 \%  $^{238}$U. Figure~\ref{Fig:DecRateComp} shows the contours in the coupling constant and HNL mass plane assuming 1 mHz and $1~\mu$Hz signal decay rates. The differences in shape, mainly at higher HNL masses, are due to a higher or lower $^{238}$U fission fraction in the different fuel types. 

Very short baseline neutrino experiments combine interesting features of being located at very close proximity to the reactors, and capable of efficiently reducing the background via pulse shape discrimination and/or segmentation. Their detection technologies are typically based on liquid and plastic/composite scintillators. An overview of the existing and planned detectors of this type can be found in Ref.~\cite{lasserre2014light}. While their designs are optimized to address long standing anomalies pointing to new mass states at  $\Delta m^2\sim 0.1~\text{eV}^2$, which effectively correspond to detector configurations operating at $L/E\sim 1~\text{m/MeV}$~\cite{PhysRevD.87.073008}, most of them could probe the decay in flight to electron-positron pairs induced by HNLs without any specific modification to their design or operations.  

%% file: chapters/Background.tex
\section{Expected rates in the Solid, JUNO and TAO detectors}

Reactor neutrino experiments are primarily designed to detect electron anti-neutrinos via the Inverse Beta Decay (IBD) reaction resulting in the detection of the positron (and its annihilation with an electron in the detector material). This prompt signal, followed by a second (delayed) signal emitted by the thermalised neutron that is captured on nuclear isotopes. Typically, these are nuclei with a high thermal neutron capture cross section, such as Lithium (SoLid) or Gadolinium (Daya Bay). However, that's not always required. For instance, JUNO and TAO identify the neutron capture on Hydrogen and Carbon. In fact, the HNL production in $\beta$-decays happens (Figure~\ref{fig:HNLdecaymode} left) inside the reactor, and it is not observed in the experiment. Therefore, the HNL signal will be identified via the observation of its decay (Figure~\ref{fig:HNLdecaymode} right), where a prompt-delay signal coincident pattern is absent in contrast to reactor $\overline{\nu}_e$ IBD interactions, as the $e^+$ and $e^-$ produced in the HNL decay are seen by the experiments as a single event. We refer to these non-coincident signals as "singles", for which dedicated background study and rejection strategy needs to be developed (see Section~\ref{s:solid}).

Even though some of the short/medium baseline reactor neutrino detectors listed in Section~\ref{sec:reactor} have a physics trigger designed to select the IBD coincident signals, they also have an electromagnetic energy threshold trigger in place that is used to either calibrate or monitor the detector stability. This trigger scheme is suitable to detect the decay products of an HNL in the mass range between 1 and 12 MeV, which is the most sensitive energy range for reactor neutrino experiments. At lower energies, the decay rate is not high enough for its detection. At higher energies, the flux becomes too small. JUNO and TAO will not trigger directly on IBD coincidences but they use a vertex fitting logic trigger~\cite{Gong:2015yvq} that is sensitivity to non-coincident signals above $\sim$300 keV. 

Three different detectors are considered in this work: i) the SoLid like detector (with size of 80 cm by 80 cm by 250 cm), ii) the TAO detector (sphere of 1.8 m diameter)~\cite{JUNO:2020ijm} and iii) the JUNO experiment (bigger version of TAO with 35.4 m diameter)~\cite{JUNO:2021vlw}. The first is next to (6~m distance) the BR2 reactor in Belgium (60 MW). The second is 30~m from one of the Taishan reactor cores (4600 MW) in China. The latter, also in China, is at a medium distance (53 km) from 8 reactor cores dispatched in two different power plants ($\sim 27$ $10^{3}$ MW). 

\begin{figure}[h!]
    \includegraphics[width=0.5\textwidth]{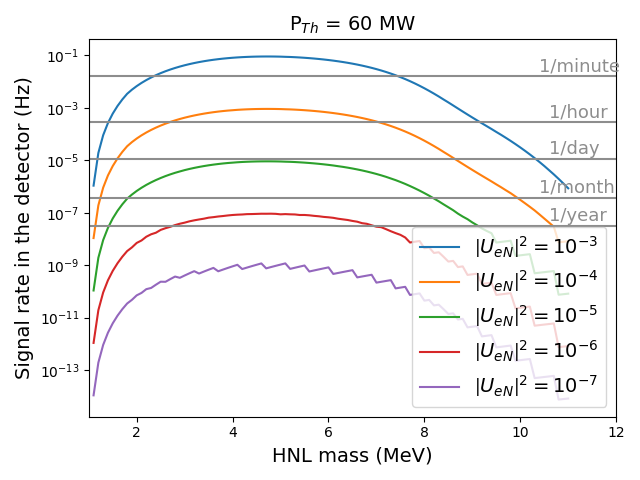}
    \includegraphics[width=0.5\textwidth]{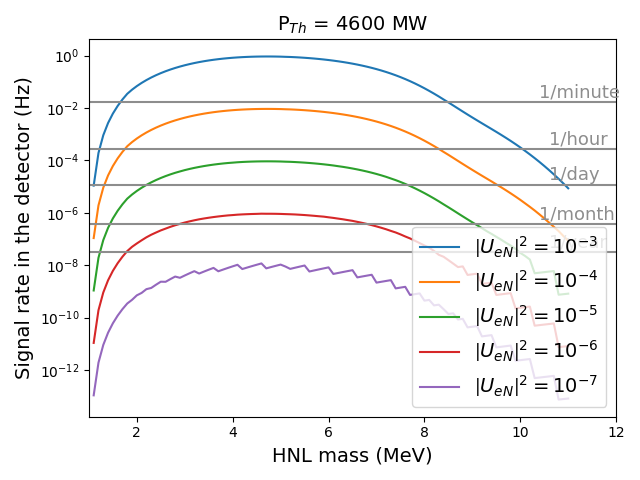}
    \caption{Rate of HNL decays in the SoLid (left) and the TAO (right) detectors.}
    \label{fig:HNLDecayRate_lines}
\end{figure}

With all the ingredients described above, the expected HNL signal rate in the various detectors can be computed. For that, the expected HNL flux is combined with its predicted decay rate and fraction of decays that will happen inside the detector. The results are shown in Figure \ref{fig:HNLDecayRate_lines} for a SoLid-like and TAO-like detectors. As a reference, some lines with recognisable rates are added shown by the horizontal lines.

\begin{figure}[h!]
    \begin{center}
    \includegraphics[width=0.65\textwidth]{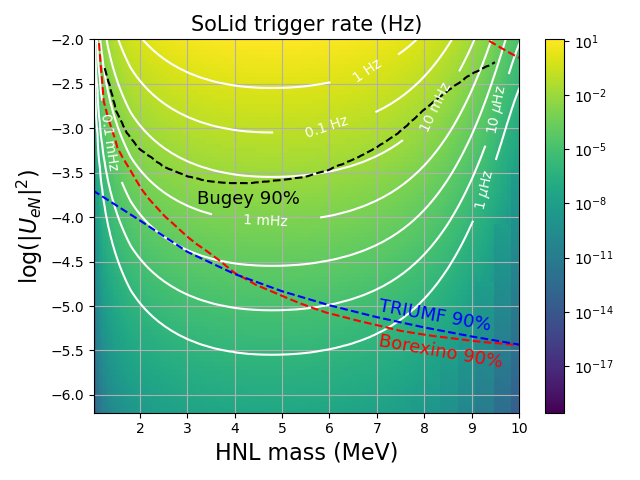}
    \end{center}
    \hspace{-0.5cm}\includegraphics[width=0.55\textwidth]{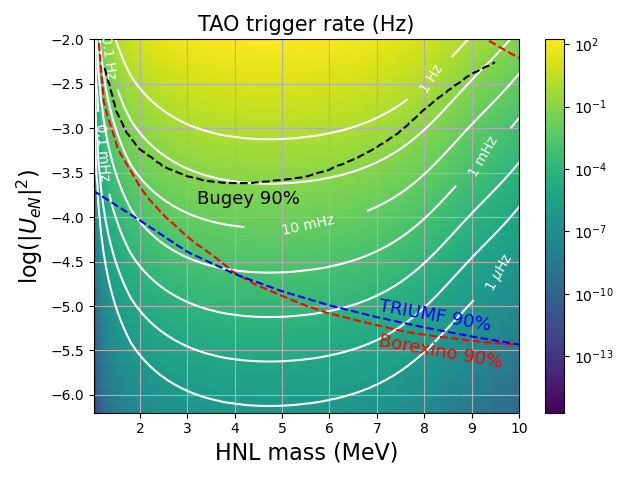}
    \includegraphics[width=0.55\textwidth]{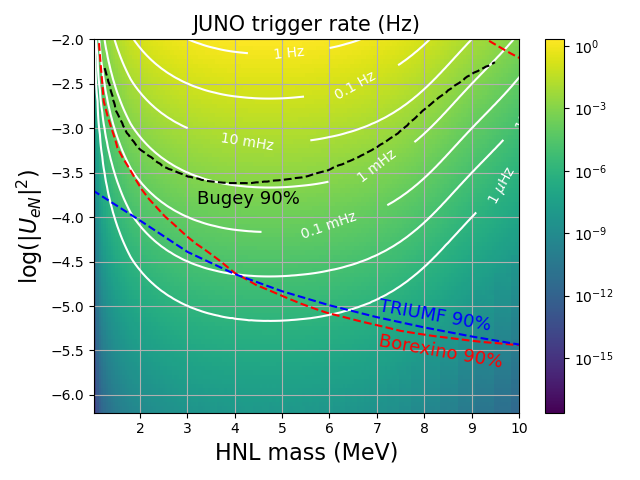}
    \caption{Expected HNL trigger rate for the SoLid detector (top), JUNO (top left) and TAO (top right) detectors. As a reference for the analyzer, the excluded regions from the Bugey~\cite{PhysRevD.52.1343}, Borexino~\cite{Borexino:2013bot} and TRIUMF~\cite{PhysRevD.46.R885} experiments are superimposed. These bounds were obtained from HNL decay searches from reactor neutrinos at Bugey and from solar neutrinos at Borexino. For TRIUMF, the contour was obtained with $\pi \rightarrow e \nu$ peak searches, not discussed here. \label{fig:trigRate}}
\end{figure}

The expected HNL signal rates in the detector can be converted into a trigger rate by using simple kinematics, together with the values provided by the experiment's design. For JUNO and TAO, the trigger efficiency for a visible energy above 1 MeV is 100\% (see Refs.~\cite{JUNO:2021vlw, JUNO:2020ijm}), while for SoLid the trigger efficiency is low for visible energies below 2 MeV and therefore this energy threshold is assumed instead. Figure \ref{fig:trigRate} shows the expected trigger rates for the three experiments as a function of the HNL mass and coupling ($m_N$,$\left|U_{eN}\right|^2$), where the previous mentioned energy thresholds are taken into account. Previous experiments have already placed exclusion limits on the parameter space investigated in this work. These limits are superimposed on the trigger rate, indicating what is the signal rate the experiment needs to extract over the background in order to achieve a competitive result. Figure \ref{fig:trigRate} shows the complementarity between the different detectors. In the case of TAO, the higher reactor power and larger size compensate for its distance compared to SoLid (5 times closer). Comparing JUNO to SoLid, its larger size and higher reactor power do not compensate for the distance to the reactor. However, the more efficient trigger capabilities allow to obtain similar constrains at low (1-2.5 MeV) and high ($>$8~MeV) HNL masses with both JUNO and SoLid.

As mentioned before, in this analysis, the signal consists in the light emitted by the electron-positron annihilation produced in the HNL decays (singles). Therefore, the HNL search will face an important background contamination, which needs to be removed for a better understanding of the signal candidates. In the context of the IBD coincidence analysis, all relevant backgrounds have already been studied extensively, but when looking for singles, additional backgrounds need to be taken into account.  

The TAO experiment will be deployed $\sim$10~m underground, so the cosmic and atmospheric background is reduced. Moreover, it is shielded by several sub-detectors to mitigate and veto the major backgrounds. The outer shielding includes: i) 1.2 m of water in the surrounding tanks, instrumented with photomultipliers and acting as a Cherenkov veto, ii) 1 m of plastic scintillator (high density polyethylene) on the top comprising the muon veto system, and iii) 10 cm of lead at the bottom. In this way, the total background rate for singles before selection, dominated by internal radioactivity, is of only $\sim$100~Hz~\cite{JUNO:2021vlw}, while the cosmogenic background after the muon veto remains below 1~Hz. In the case of JUNO, the cleanliness of the environment further reduces the internal radioactivity compared to TAO. Moreover, in JUNO and TAO the reconstruction efficiency is close to 100\% in the energy range considered here, with an excellent energy resolution of $\sim$3 and 1\%, respectively. This will allow these detectors to use the energy spectrum to efficiently separate the signal and background via a pulse shape discrimination techniques, as shown for the analysis of solar neutrino elastic scatterings in the 2-16~MeV energy range, where a signal to noise ratio (S/N) of $O$(1) is achieved (see Table 4 in~\cite{JUNO:2020hqc}). The reactor off trigger rate in SoLid before any pre-selection applied is of the order of 2~kHz (section 3.2 in~\cite{SoLid:2020cen}). As it initially has a larger background, we take the SoLid detector as an extreme case to evaluate the background mitigation impact in Section~\ref{s:solid}. 

\section{Background mitigation in SoLid}
\label{s:solid}

\subsection{Intrinsic radioactivity}

A huge background activity originates from the natural radioactivity  of the materials used in detector construction itself, and surrounding it. Among them, $^{238}$U, $^{232}$Th, $^{40}$K are the dominating background sources found in all materials. The radioactive isotope $^{208}$Tl is also a common background in experiments looking for rare events. A prompt emission of a $\beta$-decay electron with a delayed $\alpha$ particle with a time delay of tens of micro seconds with $\beta$ emission is an specific signal of the $^{238}$U chain. It is followed by the so-called BiPo decay, which consists in the decay of $^{214}$Bi to $^{214}$Po. The $\beta$-decays of the $^{214}$Bi generate a "single" signal that closely resembles that of the HNL decay. However, an $\alpha$-decay occurs subsequently, transforming the daughter nucleus, $^{214}$Po, into stable $^{210}$Pb. Discrimination of this BiPo induced background is achieved by identifying the coincidence between the prompt Bi ($\beta$-induced) and delayed Po ($\alpha$-induced) scintillation signals.
  
\subsection{Cosmics and atmospheric muons}
Particles of cosmogenic origin (specially atmospheric muons) cause the more important backgrounds for the HNL analysis. Fast neutrons, created by spallation processes when the muon collides in the atmosphere, leading to a "single" signal that mimics the searched for HNL signal. These muons can also spallate neutrons from the material that surrounds the detector when they reach it, adding an additional background. In addition, if the muon merely clips the edge of the detector, the signal might be poorly identified and resemble to that of the HNL. The muon can also be stopped in the detector and decay, resulting in yet an additional background for the HNL analysis.

\subsection{Missing backgrounds} 
The behaviour of the missing background can be inferred from the reactor-off signal that remains after subtracting the backgrounds that are already simulated (i.e. $^{214}$Bi-$^{214}$Po, cosmic muons and cosmic-induced neutrons). Also the natural radioactive elements are ingrained in the concrete and construction materials. To investigate the different reactor gamma sources, dedicated studies and data treatments need to be carried out using the spatial distribution and energy spectrum of the different signals.

 \subsection{Mitigation strategies}
The data selection criteria using direct cuts on different variables will greatly reduce the background rate, but a more effective mitigation strategy will be achieved using of the Machine Learning (ML) techniques.  Out of the many available ML methods, one of the most commonly used are boosted decision trees (BDT) due to the easiness in understanding the algorithm and the impact of the input parameters. A previous study conducted using MiniBooNE Monte Carlo samples showed that BDT has a better signal identification performance than Artificial Neural Networks (ANN) \cite{ROE2005577}. This algorithm is implemented within the TMVA package of ROOT \cite{Hocker:2007ht}. A gradient BDT is also available in the listed methods of this package. The BDT combines several discriminating variables into one final discriminator using correlations between them. This gives a better signal to background separation than applying a cut on individual variables at a time. The main challenge in using BDTs is that any mismatch in the distributions between the real data and the Monte Carlo (MC) due to a miss-modeling of the background in the MC will be reflected in the training output, which can mimic that of signal events. Hence the input parameters should be correctly modeled in the Monte-Carlo. A comparison between background selected real data sample and corresponding MC backgrounds  before training BDT will help to solve this issue. Another issue with BDTs is that they will learn statistical fluctuations by heart and are therefore very easy to overtrain.

A preliminary study done by the SoLid collaboration shows that a first pre-selection can already bring the reactor off rate to $\sim$100~Hz (same as TAO)~\cite{Verstraeten:2021bcw}. The S/N can be further improved by a factor of $\sim$200 using this kind of ML data analysis techniques feed with simple input features for each event signal cluster: its position, the reconstructed energy and the number of signal clusters being the most relevant ones.

\section{Results and discussion} 

\begin{figure}[!h]
    \centering
    \includegraphics[width=0.48\textwidth]{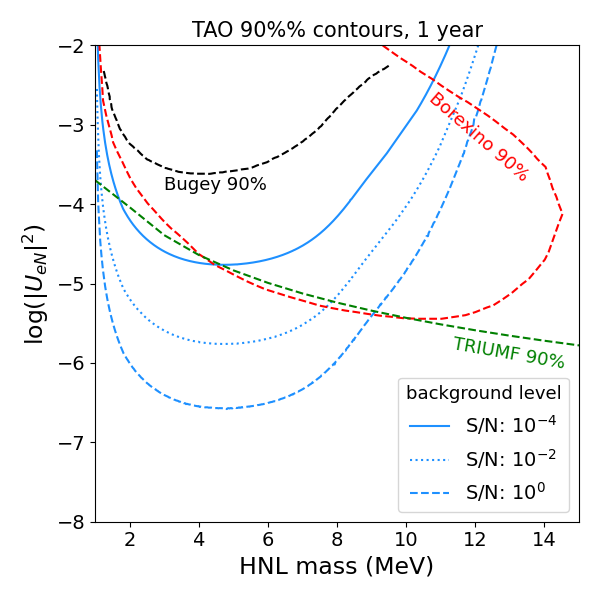}
    \includegraphics[width=0.48\textwidth]{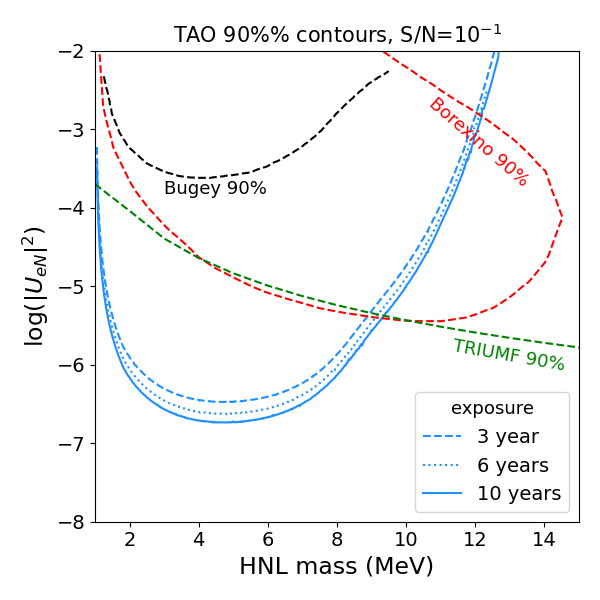}
    \caption{TAO sensitivity contours at 90\% CL. On the left, for 1 year of data taking (similar exposure than SoLid Phase 1) and three different S/N values. On the right, for 3, 6 and 10 years of exposure, with a realistic S/N = 10$^{-1}$ (conservative value compared to solar analysis).}
    \label{fig:tao}
\end{figure}

\begin{figure}[h!]
    \centering
    \includegraphics[width=\textwidth,height=7cm]{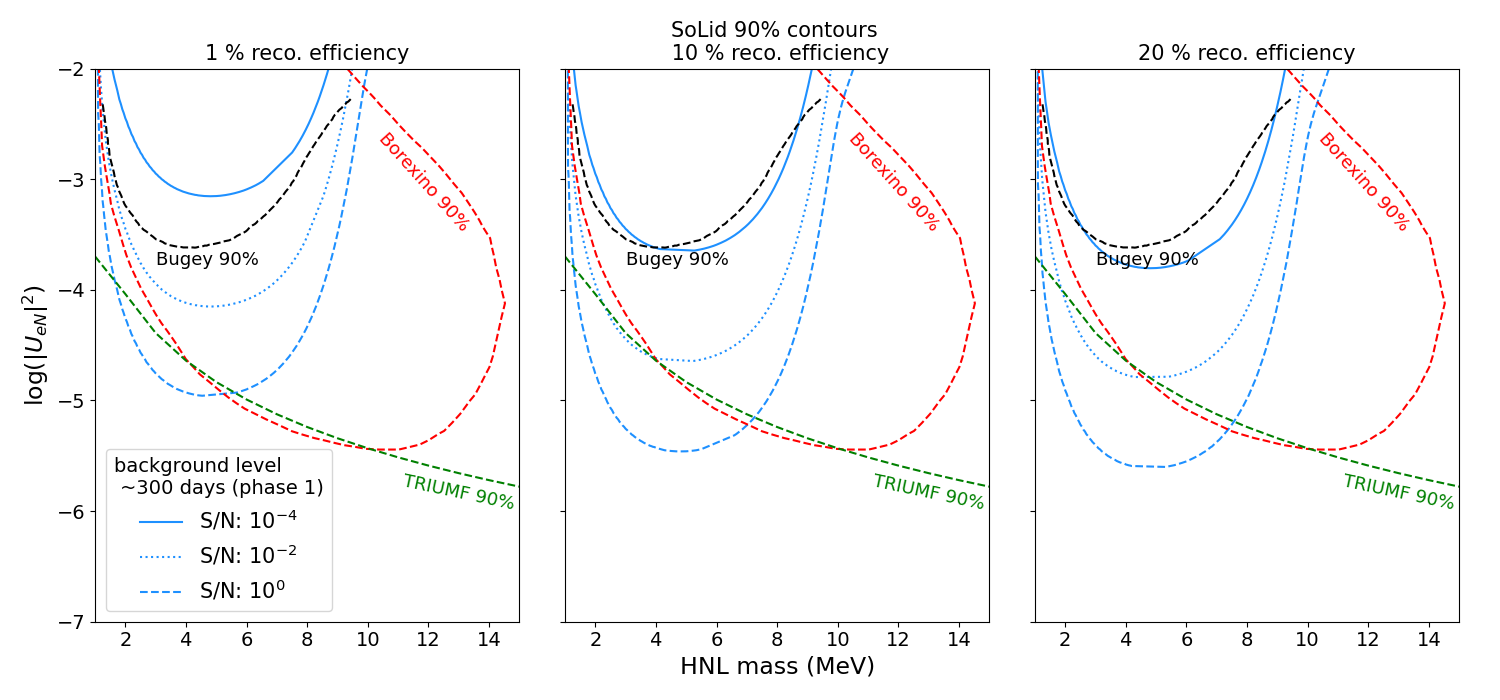}
    \caption{Expected limits at 90\% CL for the SoLid detector (Phase 1, $\sim$300 days) in the accessible HNL mass region for different signal to noise regimes and reconstruction efficiencies, the latter ranging from 1 to 20\%. \label{fig:solid}}
\end{figure}

\begin{figure}[h!]
    \centering
    \includegraphics[width=0.55\textwidth]{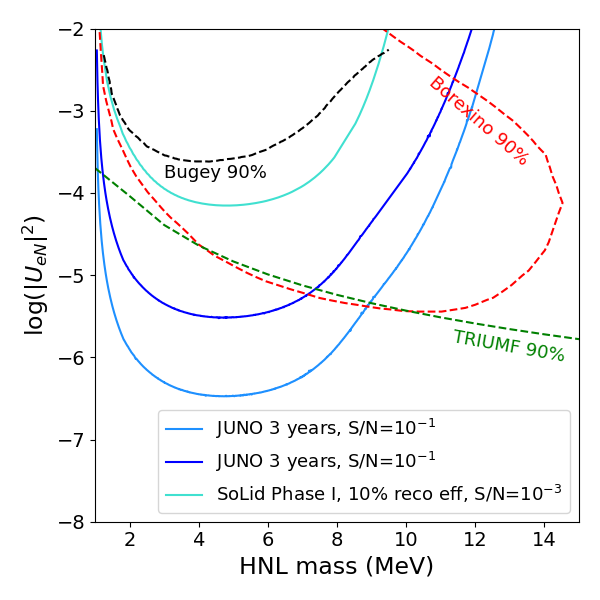}
    \caption{Expected limits (90\% C.L.) for the SoLid (cyan), JUNO (light blue) and TAO (dark blue) detectors. A reconstruction efficiency of 10\% is assumed for the SoLid experiment, while JUNO and TAO reconstruct all of the events in the reactor energy range. TAO will take data for at least three years, which is the exposure considered here for both TAO and JUNO, while for SoLid the data taking period of Phase 1 is used. \label{fig:all}}
\end{figure}

The final sensitivity contours in the ($m_{N}$,$|U_{eN}|^2$) parameter space will depend on the expected background rate, the total exposure, and the reconstruction efficiency. Awaiting a final official analysis, we have produced sensitivity contours at 90\% confidence level (CL) using various hypotheses for these parameters. For TAO (Figure~\ref{fig:tao}), we have considered different data taking periods (1, 3, 6 and 10 years), while for SoLid (Figure~\ref{fig:solid}) we have considered Phase 1 duration ($\sim$300 days). Figure~\ref{fig:all} puts together the expected contours obtained for the SoLid, TAO and JUNO detectors, taking a relatively realistic signal to noise ratio (S/N) for each of them: we assume that JUNO and TAO can reach a S/N of 10$^{-1}$, while for SoLid the best S/N can be 10$^{-3}$. The results are compared to the limits set by Bugey~\cite{PhysRevD.52.1343} (black), Borexino~\cite{Borexino:2013bot} (red) and Triumf~\cite{PhysRevD.46.R885} (green) experiments. We can see the complementary energy range probed by the reactor neutrino experiments compared to accelerator experiments like Triumf and the solar neutrino data of Borexino. If a S/N of $O$(1) is achieved in TAO (probed possible with the techniques used in~\cite{JUNO:2020hqc}), it may set the best HNL laboratory limits up to now between 1 and 8-9 GeV with only 1 year of data taking. On the other hand, we note that even if the expected signal rates are initially larger in SoLid than in JUNO (see Figure~\ref{fig:trigRate}), after taking into account the reconstruction efficiency, together with the larger exposure and better S/N achieved in JUNO with respect to SoLid, the former has a better performance at the end.


%% file: chapters/Conclusion.tex
In this work, we studied the capabilities of short and medium baseline reactor experiments to search for HNLs with masses in the MeV range and constrain their properties. Our results highlight the feasibility and discovery potential of such searches. As an example, a sensitivity estimate has been made for three reactor experiments: Solid, JUNO and TAO, showing promising capabilities. We encourage the short baseline neutrino community to use their datasets to conduct a search for MeV HNL at reactor experiments, which can lead to the best laboratory limits for HNL masses in the range 2-9 MeV.

\section*{Acknowledgements}
We acknowledge the support from the Belgian funding agencies F.R.S.-FNRS and FWO, in particular via the “Excellence of Science – EOS” – be.h project n. 30820817 and the IISN project 4.4501.17 (40008230). We also thank the SoLid and JUNO collaborations for supporting this study and for their valuable inputs.